\documentclass[aps,pra,showpacs,preprintnumbers,amssymb,amsfonts,floatfix,a4paper,twoside,twocolumn]{revtex4-1}
\usepackage{amsmath,amssymb,epsfig}
\usepackage{color}
\usepackage{float}
\usepackage{enumerate}
\usepackage{hyperref}
\usepackage[utf8]{inputenc}
\usepackage{array}
\usepackage{hhline}

\newcommand{\ket}[1]{ \left|#1\right>}

\newcommand{\outerp}[2]{ \left|#1\right>\left<#2\right|}

\usepackage{soul}
\usepackage{graphicx}
\begin{document}

\title{{\Large{\bf Detection of ultracold molecules using an optical cavity}}}

\author{Rahul Sawant}
\email{rahul.v.sawant@durham.ac.uk}
\affiliation{%
	Raman Research Institute, C. V. Raman Avenue, Sadashivanagar, Bangalore 560080, India}
\affiliation{%
	Joint Quantum Centre (JQC) Durham-Newcastle, Department of Physics,
	Durham University, South Road, Durham DH1 3LE, United Kingdom}
\author{Olivier Dulieu}
\affiliation{%
	Laboratoire Aim\'e Cotton, CNRS, Universit\'e Paris-Sud, ENS Paris-Saclay, Universit\'e Paris-Saclay, Orsay Cedex 91405, France}
\author{S. A. Rangwala}
\email{sarangwala@rri.res.in}
\affiliation{%
	Raman Research Institute, C. V. Raman Avenue, Sadashivanagar, Bangalore 560080, India}

\begin{abstract}
We theoretically study non-destructive detection of ultracold molecules, using a Fabry-Perot cavity. Specifically, we consider vacuum Rabi splitting where we demonstrate the use of collective strong coupling for detection of molecules with many participating energy levels. We also consider electromagnetically induced transparency and transient response of light for the molecules interacting with a Fabry-Perot cavity mode, as a mean for non-destructive detection. We identify the parameters that are required for the detection of molecules in the cavity electromagnetically induced transparency configuration. The theoretical analysis for these processes is parametrized with realistic values of both, the molecule and the cavity. For each process, we quantify the state occupancy of the molecules interacting with the cavity and determine to what extent the population does not change during a detection cycle.
\end{abstract}
\maketitle

\section{Introduction}\label{mol_detection}
There are a number of motivations for creating ultracold molecules, such as quantum computation~\cite{demille_quantum_2002} and quantum simulation~\cite{micheli_toolbox_2006} with polar molecules, exotic phases of matter with long range interaction between particles~\cite{pupillo_condensed_2009}, controlling chemical reactions~\cite{krems_cold_2008}, the study of few body dynamics of identical particles~\cite{quemener_ultracold_2005}, etc. In addition, trapping of polar molecules will enable a study of potentials of the form $1/r^6$ (rotating dipoles), $1/r^3$ (static dipoles), and $1/r^2$ (ion + static dipole) when co-trapped with ions. Here $r$  is the distance between the interacting particles. 

To meet these and other goals, great effort is geared towards developing techniques to create long-lived molecules at ultracold temperatures~\cite{deiglmayr_formation_2008,lang_ultracold_2008,danzl_ultracold_2010,aikawa_coherent_2010,takekoshi_ultracold_2014,molony_creation_2014,park_ultracold_2015,guo_creation_2016,stuhl_magneto-optical_2008,barry_magneto-optical_2014,steinecker_improved_2016}, particularly in their rovibrational ground state~\cite{deiglmayr_formation_2008,lang_ultracold_2008,danzl_ultracold_2010,aikawa_coherent_2010,takekoshi_ultracold_2014,molony_creation_2014,park_ultracold_2015,guo_creation_2016}, and to detect them efficiently. In most experiments, detection is destructive and requires multiple cycles of system preparation and detection~\cite{stwalley_photoassociation_1999,gabbanini_cold_2000,fatemi_ultracold_2002,herbig_preparation_2003,winkler_repulsively_2006,lang_ultracold_2008,zirbel_heteronuclear_2008}. The techniques which are used to detect molecules in most experiments rely on (i) the ionization of the molecules and the detection of the resulting ion on an ion detector~\cite{stwalley_photoassociation_1999,gabbanini_cold_2000,fatemi_ultracold_2002}, or (ii) the conversion of the molecules back to atoms and the detection of the atoms~\cite{herbig_preparation_2003,winkler_repulsively_2006,lang_ultracold_2008,zirbel_heteronuclear_2008}, or (iii) photon shot-noise limited absorption imaging on a strong but open bound-bound molecular transition~\cite{wang_direct_2010}. All these detection techniques result in destruction of the molecules at the end of the detection cycle. The development of a non-destructive technique to detect state-selected ultracold molecules would represent a major accomplishment for the study of ultracold molecules. This would enable repeated interrogation of the same molecular ensemble, building experimental statistics, and the tracking of the dynamics of the molecular ensemble.	

The goal of the present article is to devise a detection technique for ultracold molecules in a non-destructive manner, through the enhancement of their interaction with an electromagntic field generated in an optical cavity. Two concave, reflecting mirrors form a Fabry-Pérot cavity, which can trap photons for a long duration. This trapping results in the confinement of electric field due to a single photon within a very small volume, which enhances the interaction time of the photon with the resonant atoms/ions/molecules placed coupled to the cavity. Such enhancement will be useful for not just ultracold molecules but also for organic dye molecules~\cite{wang_organic}.

Unlike few-level atoms, detection of molecules using resonant light is not straightforward due to their large number of vibrational and rotational levels.
When a ground-state molecule prepared in a particular rovibrational level absorbs a photon to populate a rovibrational level of an electronic excited state, there are multiple accessible ground-state rovibrational levels to which the molecule can relax by a single spontaneous emission decay. If a rovibrational level other than the initial level is populated due to emission the molecule is lost for further imaging by the light, which is resonant with the initial rovibrational transition. However it may happen that there are a few molecules that can be detected using florescence imaging~\cite{stuhl_magneto-optical_2008} if the excited level primarily decays to a small number of ground-state rovibrational levels. However, such molecular species are seldom and cooling them to ultracold temperatures requires elaborate techniques~\cite{barry_magneto-optical_2014,steinecker_improved_2016}.

Here, we explore the possibility of detecting ultracold molecules through dispersive effects rather than absorption processes. As these dispersive effects are usually weaker than absorptive effects, we use a cavity to enhance them. We discuss various advantages and disadvantages of detecting molecules using a cavity. 
Recently, a nondestructive detection technique using Rydberg atoms was proposed for ultracold polar molecules~\cite{zeppenfeld_nondestructive_2017}. However, the introduction of other interacting species, similar to~\cite{zeppenfeld_nondestructive_2017} might result in the loss of ultracold molecules due to inelastic processes.  

Dispersive effects have been used previously to detect two-level atoms non-destructively by detecting changes in phase of light propagating through the atomic cloud~\cite{savalli_phase,lye_phase}. However, for thin atomic clouds optimized detection through phase change has an exactly same signal-to-noise ratio as optimized detection through absorption of light, if the amount of light absorbed is kept same for both the cases~\cite{lye_nondestructiveBEC}. Lye et.al~\cite{lye_nondestructiveBEC} also showed that the use of an optical cavity greatly enhances the signal-to-noise ratio for detection through phase shift. However, unlike us, they did not explore the collective effect of atoms on the cavity mode. In this article, we focus on cavity-based, non-destructive detection techniques for molecules, atoms, and ions possessing multiple levels, which use light as a measuring medium.
Assuming that an ensemble of ultracold molecules in a single quantum state is prepared, we exploit in our proposal (i) the collective strong coupling to a cavity and the corresponding vacuum Rabi splitting (VRS)~\cite{tavis_exact_1968,raizen_normal-mode_1989,thompson_observation_1992,hernandez_vacuum_2007,ray_temperature_2013} and (ii) the electromagnetically induced transparency (EIT)~\cite{bollerharris1991,fleischhauer_electromagnetically_2005,AHUFINGER2002159} for the cavity mode. The effectiveness of these options as cavity-based non-destructive mechanisms and the requirements for being practically implemented are discussed in detail. 

\section{Molecular transitions for detection}\label{molecule_levels}
There are numerous experiments which have been successful in creating molecules in the rovibrational ground state~\cite{deiglmayr_formation_2008,lang_ultracold_2008,danzl_ultracold_2010,aikawa_coherent_2010,takekoshi_ultracold_2014,molony_creation_2014,park_ultracold_2015,guo_creation_2016}. For specificity and feasibility of molecular detection in a cavity, we choose the example of ultracold Rb$_2$ molecules. We assume that the Rb$_2$ molecules are all populated in the lowest rotational and vibrational level ($\nu_\text{g} = 0, J_\text{g} = 0$) of the electronic singlet ground state $X^1\Sigma_\text{g}^+$. For all the calculations in this article the $\nu_\text{e} = 1$ and $J_\text{e} =1$ level of the electronic excited state $B^1\Pi_\text{u}$ forms the excited state for all optical transitions. For the sake of the present investigation, the $B^1\Pi_\text{u}$ state is considered as isolated, i.e. not coupled by spin-orbit interaction to other neighboring electronic states.
The relevant parameters for the transitions from this excited state are shown in Table~\ref{trans_table}. 

The above mentioned transition dipole moments are calculated using $ S_{JJ'}|\int \psi^*_{\nu_g}(R) \times d(R) \times \psi_{\nu_e}(R) \ dR|^2$~\cite{s._exploring_2015}, where $\psi_{\nu_g(\nu_e)}(R)$ is the wavefunction of the vibrational level $\nu_g(\nu_e)$ as a function of distance between the atomic cores $R$ calculated using available potential energy curves~\cite{lozeille2006,allouche_transition_2012}, $d(R)$ is $R$ dependent dipole moment for the electronic transition~\cite{beuc2007,allouche_transition_2012}, and $S_{JJ'}$ is the H\"{o}nl-London factor for the rotational levels~\cite{bernath_2005}. The wave functions, $\psi_{\nu}(R)$ are calculated using the LEVEL code~\cite{le_roy_level_2017,s._exploring_2015}, which is based on the Cooley-Numerov method. 

\begin{table}[H]
	\caption{Parameters for molecular transitions considered in this article, with $\nu_\text{e} = 1$ and $J_\text{e} =1$ level of the state $B^1\Pi_\text{u}$ as the excited state~\cite{lozeille2006,allouche_transition_2012,beuc2007}.}
	\begin{tabular}{ |m{5.45cm} | m{2.9cm}|}
		\hhline{|=|=|}
		\vspace{0.2cm}
		Transition dipole moment with $\nu_\text{g} = 0, J_\text{g} = 0$ level of $X^1\Sigma_\text{g}^+$ state & $4.8\times10^{-29}$ C$\cdot$m (14.2 Debye)\\
		\hline
		\vspace{0.1cm}
		Transition dipole moment with $\nu_\text{g} = 1, J_\text{g} = 0$ level of $X^1\Sigma_\text{g}^+$ state & $5.1\times10^{-29}$ C$\cdot$m (15.3 Debye)\\
		\hline
		\vspace{0.1cm}
		Decay rate to $\nu_\text{g} = 0, J_\text{g} = 0$ level of $X^1\Sigma_\text{g}^+$ state & 401.5 kHz\\
		\hline
		\vspace{0.1cm}
		Decay rate to $\nu_\text{g} = 1, J_\text{g} = 0$ level of $X^1\Sigma_\text{g}^+$ state & 456.6 kHz\\
		\hline
		\vspace{0.1cm}
		Total decay rate to all levels of $X^1\Sigma_\text{g}^+$ state & 6.44 MHz\\
		\hhline{|=|=|}
	\end{tabular}
	\label{trans_table}
\end{table}

In the following sections, the described processes could be considered  for atoms, ions or molecules. Therefore, we will use the word ``atom" for the sake of simplicity, unless otherwise stated when features specific to molecules will be invoked.
\section{Molecule detection with VRS}\label{VRSsec}

\begin{figure}[!t]
	\centering
	\includegraphics[width=8.5cm]{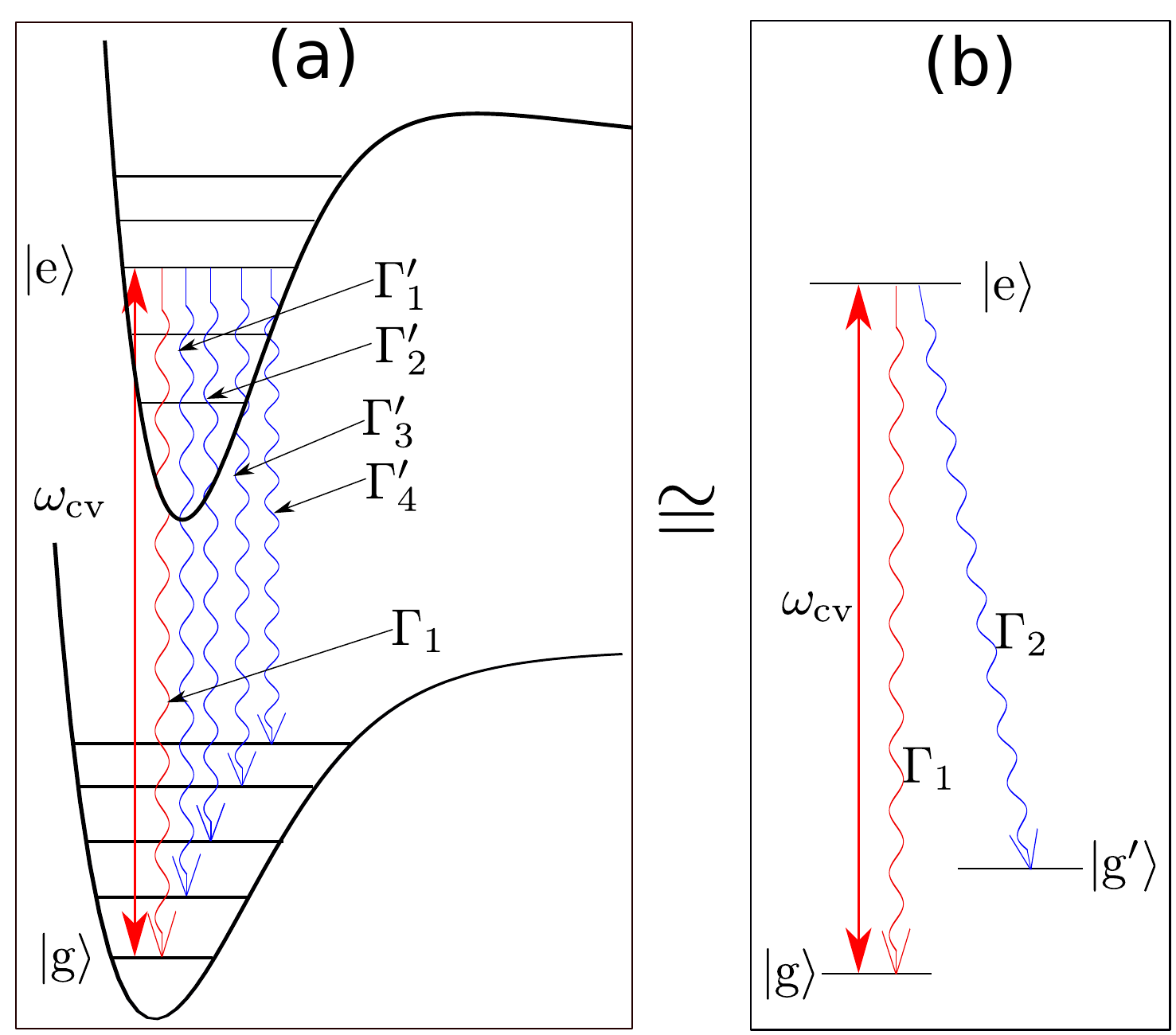}
	\caption{Schematic energy level diagram showing the equivalence of (a) the molecular levels and (b) a 3-level atomic system for the purpose of detection using VRS (see text for details). The red straight-lined arrow denotes the coupling of the $\ket{\text{g}}$ and $\ket{\text{e}}$ levels by the cavity photon with energy $\hbar \omega_\text{cv}$, and the wavy-lined arrows denote spontaneous emission processes labeled with their relevant rates. In panel (b) $\Gamma_2 = \Gamma'_1+\Gamma'_2+\Gamma'_3+\Gamma'_4$ is the total rate of spontaneous emission to ground-state vibrational levels other than $\ket{\text{g}}$ and $\omega_\text{cv}$. }
	\label{molecule_vrs}
\end{figure}

When multiple atoms are coupled to a cavity such that the atom-cavity system is in the collective strong coupling regime, the empty cavity single peak of transmission for a probe light beam through the cavity splits into two non-degenerate transmission peaks. This splitting is called vacuum Rabi splitting~\cite{tavis_exact_1968,raizen_normal-mode_1989,thompson_observation_1992,hernandez_vacuum_2007,ray_temperature_2013}. The frequency separation between the two peaks is equal to $2g_0\sqrt{N_c}$, where $g_0$ is coupling strength of a single atom with the cavity and $N_c$ is the effective number of atoms coupled to the cavity. The measured splitting between the two VRS peaks informs on the number of atoms coupled to the cavity. This splitting is clearly observed for a two-level atom, where the spontaneous radiative decay of the excited level $\ket{\text{e}}$ back to the ground-state level $\ket{\text{g}}$ does not uncouple the atom-cavity system. 
When the VRS is measured on an atom with an additional level $\ket{\text{g'}}$, the atoms get rapidly pumped in the third state leading to a collapse of the VRS signal. In such a simple level scheme the VRS signal can be recovered by the addition of another laser to offset the to ensure the repumping of the $\ket{\text{g'}}$ population into the $\ket{\text{g}}$ level.~\cite{ray_temperature_2013}. However this is not possible in molecules with large number of loss channels. Having such a repumping laser is not practical for most molecular species due the large number of rovibrational levels in the ground state of the molecules, which prevents the realization of a closed optical transition~\footnote{There are few recent works which do use such a repumping scheme for the purpose of laser cooling of specific molecules~\cite{barry_magneto-optical_2014,steinecker_improved_2016}. However, after scattering of few thousand photons the molecules are lost for further laser cooling as levels not accessible any more by the laser are populated.}. Therefore, the question is, whether molecules can at all be detected using VRS, and if so under which conditions? A plausible strategy is to check if the VRS measurement is much faster than the photon absorption rate of the cavity-coupled species. In this case, the detection could be achieved without significantly changing the number of molecules coupled to the cavity. To evaluate the validity of this hypothesis, we consider an equivalent case of an idealized 3-level atom with two ground-state levels coupled to a cavity where the cavity is resonant with one of the transitions, and there is no repump light present as shown in Fig.~\ref{molecule_vrs}. 
The Hamiltonian for an ensemble of N three-level atoms (as shown in Fig.~\ref{molecule_vrs}), in a frame rotating at the frequency of the probe laser ($\omega_\text{p}$) probing the atom-cavity system, the Hamiltonian, assuming stationary atoms is,
\begin{equation}
\hat{H}= \hbar \sum^{N}_{j = 1} \left[-\Delta_\text{pa} \hat{\sigma}^{j}_\text{ee} + g_j  (\hat{a}^{\dagger} \hat{\sigma}^{j}_\text{ge}+ \hat{a} \hat{\sigma}^{j}_\text{eg}) \right]. 
\label{VRS_ham}
\end{equation}
Here, $\hat{\sigma}^{j}_{mn} = (\outerp{m}{n})^{j}$ denotes the atomic operators for $j^{\text{th}}$ atom, $\Delta_\text{pa}= \omega_\text{p}-(\omega_\text{e} - \omega_\text{g})$ is the probe laser detuning from the $\ket{\text{g}}\leftrightarrow\ket{\text{e}}$ transition, $g_j = g_0 f(x_j,y_j,z_j)$ is coupling of $j^\text{th}$ atom with the cavity mode with $g_0=-\mu_\text{ge}\sqrt{\omega_\text{cv} /(2 \hbar \epsilon_0 V)}$ as the maximum atom-cavity coupling, $\mu_\text{ge}$ is the electric dipole moment for the transition coupled to the cavity, $\omega_\text{cv}/(2\pi)$ is the resonance frequency of the cavity, $V$ is the volume of the cavity mode, $f(x,y,z)$ is the mode function of the cavity, $\hat{a}$ and $\hat{a}^{\dagger}$ are the photon annihilation and creation operators for the cavity field.
The evolution equation for the expectation value of an operator $\hat{X}$ can be evaluated using the Heisenberg equation,
\begin{equation}
\frac{d\langle\hat{X}\rangle}{dt} = \frac{i}{\hbar} \langle [\hat{H},\hat{X}]\rangle,
\label{hisen}
\end{equation}
where, $[\hat{H},\hat{X}]$ is the commutator of $\hat{X}$ with $\hat{H}$. 

\begin{figure}[!t]
	\centering
	\includegraphics[width=8.5cm]{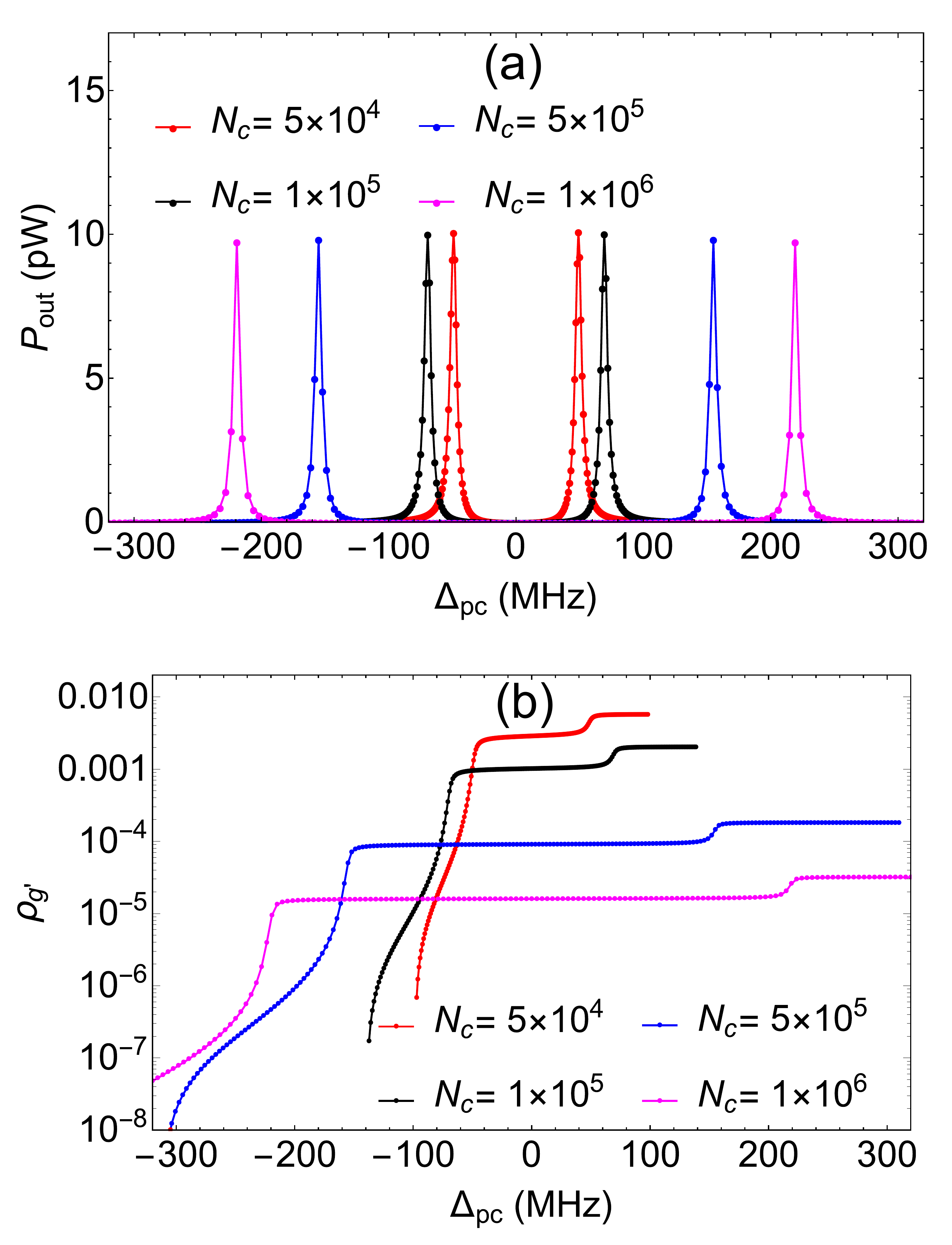}
	\caption{(a) The output power ($P_\text{out}$) of the cavity-atom system as a function of the detuning of the probe laser from the cavity frequency showing VRS for various atom numbers. (b) Probability for the atoms to end into the dark ground state $\ket{\text{g}'}$ in a single sweep across the atom-cavity resonances. At the start of the scan, all atoms are assumed to be in state $\ket{\text{g}}$.} 
	\label{3level_VRS_scan}
\end{figure}

For the atom-cavity system defined above, the evolution equations for the atomic states and the cavity field, after including spontaneous emission rates, 
and assuming the cavity field to be classical denoted by a coherent state $\ket{\alpha}$ result in the set of coupled differential equations~\cite{scully_quantum_1997,walls_quantum_2008,sawant_interactions_2017},
\begin{subequations}\label{3levelvrseq}
	\begin{align}
	\frac{d\alpha(t)}{dt} &=  -\left(\kappa_\text{t} - i\Delta_\text{pc}\right) \alpha(t)  - ig_0N_\text{c}\rho_\text{ge}(t) - \eta \\
	\frac{d\rho_\text{ge}(t)}{dt} &= -\left\{\frac{\Gamma_\text{t}}{2} - i \Delta_\text{pa} \right\} \rho_\text{ge}(t)  + i g_0 \alpha(t) (\rho_\text{e}(t) - \rho_\text{g}(t))   \\
	\frac{d\rho_\text{e}(t)}{dt} &=  -\Gamma_\text{t} \rho_\text{e}(t)+ i g_0\left\{ \alpha^{*}(t) \rho_\text{ge}(t) -  \alpha(t) \rho_\text{ge}^{*}(t) \right\}\\
	\frac{d\rho_\text{g}(t)}{dt} &=  \Gamma_1 \rho_\text{e}(t)- ig_0\left\{  \alpha^{*}(t) \rho_\text{ge}(t) -  \alpha(t) \rho_\text{ge}^{*}(t) \right\}\\
	\frac{d\rho_{\text{g}'}(t)}{dt} &=  \Gamma_2 \rho_\text{e}(t).
	\end{align}
\end{subequations}    
Here, $\rho$'s with single letter as subscript are the populations of the levels and the ones with two letter subscripts are coherences between levels, $\Delta_\text{pc}= \omega_\text{p}-\omega_\text{cv}$, $\Gamma_1$ and $\Gamma_2$ are the decay rates of the excited state $\ket{\text{e}}$ to the ground states $\ket{\text{g}}$ and $\ket{\text{g}'}$ respectively, $\eta$ is the rate at which classical light is injected into the cavity from the incident probe light, $\kappa_\text{t}$ is the decay rate of the cavity field, and $\Gamma_\text{t}= \Gamma_1+\Gamma_2$ is the total decay rate of the excited state. $\ket{\text{g}'}$ represents all the dark ground states of molecules as can be seen from Fig.~\ref{molecule_vrs}, the major role of this dark ground state is to bleach the atoms from the transition interacting with the cavity so it is irrelevant whether it is just one dark ground state or many, for calculations in this section. For computational ease, in deriving the Eqns.~\ref{3levelvrseq} we assumed effectively $N_\text{c}$ atoms couple to the cavity with equal strength $g_0$. $N_\text{c}$ can be obtained by computing the overlap of the cavity mode function, $f(x,y,z)$ and the atomic density profile~\cite{albert_collective_2012,sawant_optical-bistability-enabled_2016,sawant_interactions_2017}. Hence we remove the subscript $j$ while going from Eqn.~\ref{VRS_ham} to Eqns.~\ref{3levelvrseq} (see~\cite{sawant_optical-bistability-enabled_2016,sawant_interactions_2017} for more details). This simplification enables solving the time-dependent differential Eqns.~\ref{3levelvrseq} by numerical integration to obtain the power of light transmitted by the atom-cavity system~\cite{sawant_optical-bistability-enabled_2016,sawant_interactions_2017}. Scanning across the atom-cavity resonance, we obtain a VRS signal, as shown in Fig.~\ref{3level_VRS_scan}. In the simulation, the detuning of the probe laser is adiabatically increased in the simulation and the corresponding change in cavity output power is monitored. Here the change is adiabatic with respect to the atomic and cavity rates. The detuning of the probe laser is scanned over 200 steps of 0.005 ms (0.1 ms in total), resulting in a scan rate of 10 kHz. The probe light power ($P_\text{in} =$ 0.23 nW) is such that the maximum output of the cavity during the detection stage is $P_\text{out} = 10$ pW, thus corresponding to $4\times10^7$ photons per second. With these parameters, the maximum number of photons available for detection at each step of the scan in Fig.~\ref{3level_VRS_scan}(a) is $10$. If the photon detection efficiency is $50\%$, this gives 50 ns delay time between two photons on average. This time duration is equal to the typical dead time of single-photon avalanche photodetectors (APD)~\cite{hadfield_single-photon_2009} making such a detection feasible. The maximum photon occupancy in the cavity mode for these parameters is three. The cavity mirror separation is taken to be $11.8$ mm and the radius of curvature of the mirrors is $10$ mm yielding a waist size of 30 $\mu$m for the cavity mode at the center of the cavity. These typical  cavity parameters are taken from the experimental work of Albert et al.~\cite{albert_collective_2012}. 
The molecular levels used here are, $\ket{\text{e}}$ is $\nu_\text{e} = 1, J_\text{e} = 0$ level of $B^1\Pi_\text{u}$ electronically excited state, $\ket{\text{g}}$ is $\nu_\text{g} = 0, J_\text{g} = 0$ level of $X^1\Sigma_\text{g}^+$ electronic ground state as explained in section~\ref{molecule_levels}. Throughout this article, we assume only the above mentioned transition couples to the cavity giving a maximum coupling constant of  $g_0/2\pi = 219.2$ kHz. $\ket{\text{g}'}$ represents all other levels of ground state $X^1\Sigma_\text{g}^+$. 
Other parameters for Fig.~\ref{3level_VRS_scan} are,
$\kappa_\text{t}/2\pi = 2.5$ MHz, loss rate from input mirror of cavity $\kappa_\text{r1} = 0.1\kappa_\text{t}$, and loss rate from output mirror of cavity $\kappa_\text{r2} = 0.8\kappa_\text{t}$.    

From Fig.~\ref{3level_VRS_scan}(b) we see that larger the number of atoms smaller the leakage into the dark ground state ($\ket{\text{g}'}$). This results due to the higher atom number shifting the resonance frequencies of the VRS peaks away from the atomic resonance and hence the probability of photon absorption for an atom reduces.  For $N_\text{c} = 5\times10^4$, $0.54\%$ of the atoms will be lost to the dark ground state per scan and for $N_\text{c} = 1\times10^6$, $3\times10^{-3}\%$ atoms will be lost. These numbers suggest that such a detection scheme is feasible. 

In order to keep the absorption and spontaneous emission of a photon by single atoms low, the photon occupancy of the cavity needs to be minimal. This results in a lower flux of photons out of the cavity, consequently reducing the detection probability. To ensure an optimal detection, the loss from cavity mirrors other than the loss due to leakage from output mirror should be minimized. This is achieved using an ultra-high reflectivity input mirror and a moderately reflective output mirror such that most of the losses from the mirrors are due to transmission rather than absorption and scattering losses. This results in a high ratio  $\kappa_\text{r2}/\kappa_\text{t}$, where $\kappa_\text{r2} $ is the loss rate from the output mirror of the cavity and $\kappa_\text{t}$ the total loss rate from the cavity mirrors. For instance, keeping the output photon flux and all other parameters identical to those of the simulation of Fig.~\ref{3level_VRS_scan}, a low ratio of $\kappa_\text{r2}/\kappa_\text{t} = 0.01$ will result in a loss of $64\%$ atoms in the case of $N_\text{c} = 5\times10^4$ and a loss of $0.3\%$ atoms in the case of $N_\text{c} = 1\times10^6$ per scan. Hence it is important to keep $\kappa_\text{r2}/\kappa_\text{t}$ as high as possible. In addition, having a larger value of $\kappa_\text{t}$ is advantageous in this case because a larger value of $\kappa_\text{t}$ means a smaller lifetime of the photon inside the cavity. This reduces its probability of being absorbed, and the steady state inside the cavity is achieved faster, enabling a faster scan. For example, if we keep the ratio mentioned above, output photon flux, and other parameters same as for the Fig.~\ref{3level_VRS_scan}, $\kappa_\text{t} = 20$ MHz will result in a loss of $0.3\%$ in the case of $N_\text{c} = 5\times10^4$ and a loss of $1.5\times10^{-3}\%$ atoms per scan in the case of $N_\text{c} = 1\times10^6$ and, $\kappa_\text{t} = 0.5$ MHz will result in loss of $2\%$ atoms in the case of $N_\text{c} = 5\times10^4$ and a loss of $12\times10^{-3}\%$ atoms per scan in the case of $N_\text{c} = 1\times10^6$. However, we cannot keep on increasing $\kappa_\text{t}$ because this will result in a very broad cavity linewidth which will engulf the VRS. This suggests that a careful choice of cavity parameters is critical for the success of the scheme.

\section{Detection through cavity EIT}  
In this section, we explore the detection of molecules interacting with a cavity using electromagnetically induced transparency~\cite{bollerharris1991,fleischhauer_electromagnetically_2005,AHUFINGER2002159}. In this phenomenon a frequency window of transparency is opened for a probe laser which would have been absorbed by an ensemble of atoms (or ions, or molecules) resonant with the probe light. This transparency is induced by another strong light beam interacting with the same 3-level ensemble. There is a steep dispersive effect, and little or no absorption for a probe light near the EIT peak~\cite{harris1992,	fleischhauer_electromagnetically_2005}. 

To start with, we will first analyze the ideal case of three-level atoms coupled to the cavity and later extend the analysis to molecules which are equivalent to leaky four-level atoms. The three levels with two ground states ($\ket{\text{g}}$ and $\ket{\text{g}'}$) along with an excited state ($\ket{\text{e}}$) form a lambda system~\cite{fleischhauer_electromagnetically_2005} as shown in Fig.~\ref{molecule_vrs}(b). In contrast to VRS, a second light field which couples the ground state ($\ket{\text{g}'}$) to excited state is required for EIT. We call it control beam which has a frequency $\omega_r$. In a frame rotating at the probe frequency ($\omega_\text{p}$), the Hamiltonian for stationary atoms for such a case is,
\begin{align}\label{EITham}
\hat{H} =& \hbar \sum^{N}_{j=1} \left[-\Delta_\text{pa}  \hat{\sigma}^{j}_\text{ee} + (-\Delta_\text{pa}+\Delta_\text{ra})  \hat{\sigma}^{j}_{\text{g}'\text{g}'} \right. \nonumber \\
 & \left. + g_j  (\hat{a}^{\dagger}  \hat{\sigma}^{j}_\text{ge}+ \hat{a} \hat{\sigma}^{j}_\text{eg}) + (\Omega^{*} \hat{\sigma}^{j}_{\text{g}'\text{e}}+ \Omega \hat{\sigma}^{j}_{\text{e}\text{g}'}) \right]. 
\end{align}
Here, $2\Omega=-\mu_{\text{g}'\text{e}} |E|/\hbar$ is the Rabi frequency for the control beam, where $\mu_{\text{g}'\text{e}}$ is the transition dipole moment for the transition $\ket{\text{g}'}\rightarrow\ket{\text{e}}$ and $E$ is the electric field amplitude of the control beam. $\Delta_\text{ra}= \omega_\text{r}-(\omega_\text{e} - \omega_{\text{g}'})$ is the control laser detuning from the $\ket{\text{g}'}\leftrightarrow\ket{\text{e}}$ transition. 
Using Eqn.~\ref{hisen} the system evolution is described by the coupled differential equations, 
\begin{subequations}\label{3leveleiteq}
	\begin{align}
	&\frac{d\alpha}{dt} = -\eta  -\alpha  (\kappa_\text{t} - i\Delta_\text{pc})-i \sum^{N}_{j=1} g_j  \rho^{j}_\text{ge}  \\
	&\frac{d\rho^{j}_\text{ge}}{dt} = (-\frac{\Gamma_\text{t}}{2}+i \Delta_\text{pa}) \rho^{j}_\text{ge} -i \alpha  g_j (\rho^{j}_\text{gg}-\rho^{j}_\text{ee})-i \rho^{j}_{\text{gg}'} \Omega  \\
	&\frac{d\rho^{j}_{\text{gg}'}}{dt} = (-\gamma_{\text{gg}'}+i \Delta_\text{pa}-i \Delta_\text{ra}) \rho^{j}_{\text{gg}'} +i \alpha  g_j \rho^{j}_{\text{eg}'}-i \rho^{j}_\text{ge} \Omega ^*  \\
	&\frac{d\rho^{j}_{\text{g}'\text{e}}}{dt} = (-\frac{\Gamma_\text{t}}{2}+i \Delta_\text{ra}) \rho^{j}_{\text{g}'\text{e}} -i \alpha  g_j\rho^{j}_{\text{g}'\text{g}} - \Omega  (\rho^{j}_{\text{g}'\text{g}'}-\rho^{j}_\text{ee})  \\
	&\frac{d\rho^{j}_\text{gg}}{dt} = \Gamma_1 \rho^{j}_\text{ee}-i \alpha ^* g_j \rho^j_\text{ge}+i \alpha  g_j \rho^j_\text{eg} \\
	&\frac{d\rho^{j}_{\text{g}'\text{g}'}}{dt} = \Gamma_2 \rho^{j}_\text{ee}+i \rho^{j}_{\text{eg}'} \Omega -i \rho^{j}_{\text{g}'\text{e}} \Omega ^* \\
	&\frac{d\rho^{j}_\text{ee}}{dt} = -\Gamma_\text{t}  \rho^{j}_\text{ee}+i \alpha ^* g_j \rho^{j}_\text{ge}-i \alpha  g_j \rho^{j}_\text{eg} - \rho^{j}_{\text{eg}'} \Omega +i \rho^{j}_{\text{g}'\text{e}} \Omega ^*. 
	\end{align}
\end{subequations}
Here, $\gamma_{\text{gg}'}$ is the decoherence rate for the coherence between the two ground states $\{\ket{\text{g}},\ket{\text{g}'}\}$.
In steady state, $\frac{d\alpha}{dt}=0$ and $\frac{d\rho_{mn}}{dt}=0, \forall(m,n)$ and Eqns.~\ref{3leveleiteq} become a set of linear equations which can be solved algebraically. Eliminating the atomic variables we get, 
\begin{align}
-\eta &  -\alpha  (\kappa_\text{t} - i\Delta_\text{pc})= i \sum^{N}_{j=1} g_j  \rho^{j}_\text{ge} \nonumber \\
&= i \alpha \sum^{N}_{j=1}\chi_j \nonumber \\
&= i \alpha \frac{2  g_0^2 N_\text{c}(\Delta_\text{ra}-\Delta_\text{pa})}{2 |\Omega|^2+(2\Delta_\text{pa}+i \Gamma_\text{t} )(\Delta_\text{ra}-\Delta_\text{pa})}\nonumber \\
&= i \alpha \chi.
\end{align}
Where, 
\begin{align}
\chi_j =\frac{2  g_j^2(\Delta_\text{ra}-\Delta_\text{pa})}{2 |\Omega|^2+(2\Delta_\text{pa}+i \Gamma_\text{t} )(\Delta_\text{ra}-\Delta_\text{pa})}
\end{align}
is the linear susceptibility of $j^{\text{th}}$ atom, $\chi$ is the total linear susceptibility, and we make use of $\sum^{N}_{j=1} g_j^2 = g_0^2N_c$~\cite{sawant_interactions_2017} as we are interested in the average effect. In deriving the above equation, we have assumed that the intra-cavity light amplitude is very small compared to other relevant parameters, i.e. $g_0|\alpha|\ll \Omega,\Gamma_\text{t}$ and hence the susceptibility $\chi$ shows linear dependence with the cavity field amplitude $\alpha$ after neglecting the small nonlinear terms. We also assume there is no decoherence between the two ground states, i.e. $\gamma_{\text{gg}'} = 0$, which is valid for a dilute gas.     
The average photon number inside the cavity can then be written as,
\begin{equation}
\bar{n} = |\alpha|^2 = \frac{\eta ^2}{(\Delta_\text{pc}-\chi_1)^2+(\kappa_\text{t}-\chi_2)^2}
\end{equation}  
where,
\begin{equation}
\chi_1 = \frac{4 g_0^2 N_\text{c} (\Delta_\text{ra}-\Delta_\text{pa}) \left(|\Omega|^2-\Delta_\text{pa} (\Delta_\text{ra}-\Delta_\text{pa})\right)}{\Gamma_\text{t}^2 (\Delta_\text{ra}-\Delta_\text{pa})^2+4 \left(|\Omega|^2-\Delta_\text{pa} (\Delta_\text{ra}-\Delta_\text{pa})\right)^2} \nonumber
\end{equation}  
and
\begin{equation}
\chi_2 = -\frac{2 \Gamma_\text{t}  g_0^2 N_\text{c} (\Delta_\text{ra}-\Delta_\text{pa})^2}{\Gamma_\text{t}^2 (\Delta_\text{ra}-\Delta_\text{pa})^2+4 \left(|\Omega|^2-\Delta_\text{pa} (\Delta_\text{ra}-\Delta_\text{pa})\right)^2} \nonumber
\end{equation}  
are real and imaginary parts of $\chi$ respectively. $\chi_1$ results in the dispersive effects and $\chi_2$ results in change of total loss rate for the cavity field. 
In the simple case where $\Delta_\text{ra} = 0$ and $\Delta_\text{pa} = \Delta_\text{pc} = \Delta$,
\begin{equation}
\chi_1 = -\frac{4 g_0^2 N_\text{c}\Delta \left(|\Omega|^2+\Delta^2\right)}{\Gamma_\text{t}^2 \Delta^2+4 \left(|\Omega|^2+\Delta^2\right)^2}\nonumber 
\end{equation}  
and
\begin{equation}
\chi_2 = -\frac{2 \Gamma_\text{t}  g_0^2 N_\text{c} \Delta_\text{pa}^2}{\Gamma_\text{t}^2 \Delta_\text{pa}^2+4 \left(|\Omega|^2+\Delta^2\right)^2}. \nonumber
\end{equation}  
In the limit of $\Delta^2 \ll |\Omega|^2, 4\frac{|\Omega|^4}{\Gamma_\text{t}^2}$, i.e. small detuning of probe laser near the EIT peak, the intracavity photon number ($\bar{n}$) reduces to,
\begin{equation}
\bar{n} = |\alpha|^2 = \frac{\eta ^2}{\kappa_\text{t}^2}\frac{d^2}{\Delta^2+d^2}.
\end{equation}
This is a Lorentzian function with full width at half maxima (FWHM), 
\begin{equation}
2d =  \frac{2|\Omega|^2 \kappa_\text{t} }{\sqrt{\Gamma_\text{t}  g_0^2 \kappa_\text{t} N_\text{c}+\left(g_0^2 N_\text{c}+|\Omega|^2\right)^2}}
\end{equation}
For $|\Omega|^2 \gg \Gamma_\text{t} \kappa_\text{t}$, the FWHM reduces to, 
\begin{equation}
2d = \frac{2\kappa_\text{t}}{(\frac{g_0^2 N_\text{c}}{|\Omega|^2}+1)}.
\label{fwhm_eit_simple}
\end{equation}
This expression yields a simple dependence of the FWHM on the atom number. The greater the atom number, the lower the linewidth of the Lorentzian. Importantly, of FWHM is linear in $N_\text{c}$ in contrast with the square root dependence of VRS, thus making the EIT method of atom number detection more sensitive. Such a linewidth narrowing for cavity transmission was predicted earlier by Lukin et al.~\cite{lukin_intracavity_1998}. It was observed for thermal atoms coupled to ring cavity~\cite{wang_cavity-linewidth_2000}, laser-cooled atoms coupled to Fabry-Perot cavity~\cite{zhang_slow_2008}, and for laser-cooled ions coupled to Fabry-Perot cavity~\cite{albert_cavity_2011}. However, for a cavity based EIT, the dependence of EIT linewidth on the number of atoms derived in this article has not been explored and requires experimental verification.
\begin{figure}[!t]
	\centering
	\includegraphics[width=8.5cm]{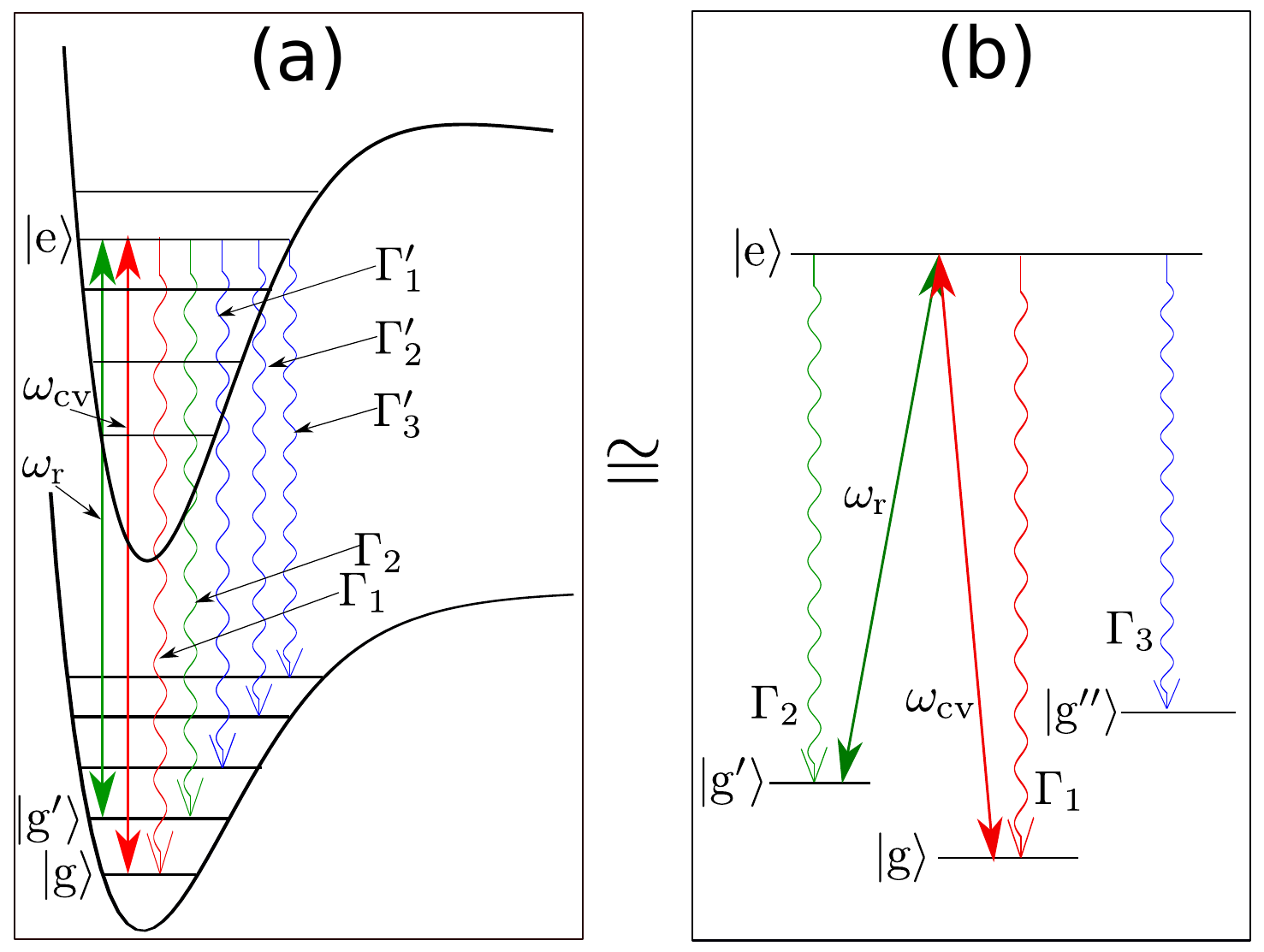}
	\caption{Representative energy level diagram showing the equivalence of (a) the molecular levels and (b) a 4-level atomic system for the purpose of detection using EIT. The straight red line denotes coupling of the cavity photon, the straight green line denote coupling of the control laser, and the wavy lines denote spontaneous emission processes with the corresponding rates shown. Here, $\Gamma_3 = \Gamma'_1+\Gamma'_2+\Gamma'_3$ is the total rate of spontaneous emission to ground vibrational levels other than $\ket{\text{g}}$ and $\ket{\text{g}'}$. }
	\label{molecule_eit}
\end{figure} 

In the above analysis, the lambda system for the atoms was a closed system, and there was no leakage to any other dark state. However, this will not be the case for molecules due to the presence of a large number of ground-state levels. To see if the Eqn.~\ref{fwhm_eit_simple} still holds for such a leaky system we solve the time-dependent partial differential equations~\ref{3leveleiteq} similar to section~\ref{VRSsec} after removing the $j$ subscript and assuming average number of atoms $N_\text{c}$ couple equally to the cavity. Here in addition to the decays in Eqns.~\ref{3leveleiteq} an additional decay $\Gamma_3$ to the dark ground state $\ket{\text{g}''}$ is also included. The total decay rate from excited state becomes $\Gamma_\text{t} = \Gamma_1 + \Gamma_2+ \Gamma_3$ (see Fig.~\ref{molecule_eit}(b)) and all other parameters remain the same. An equivalence diagram between molecular levels and atomic levels is shown in the Fig.~\ref{molecule_eit}.  
\begin{figure}[!b]
	\centering
	\includegraphics[width=7cm]{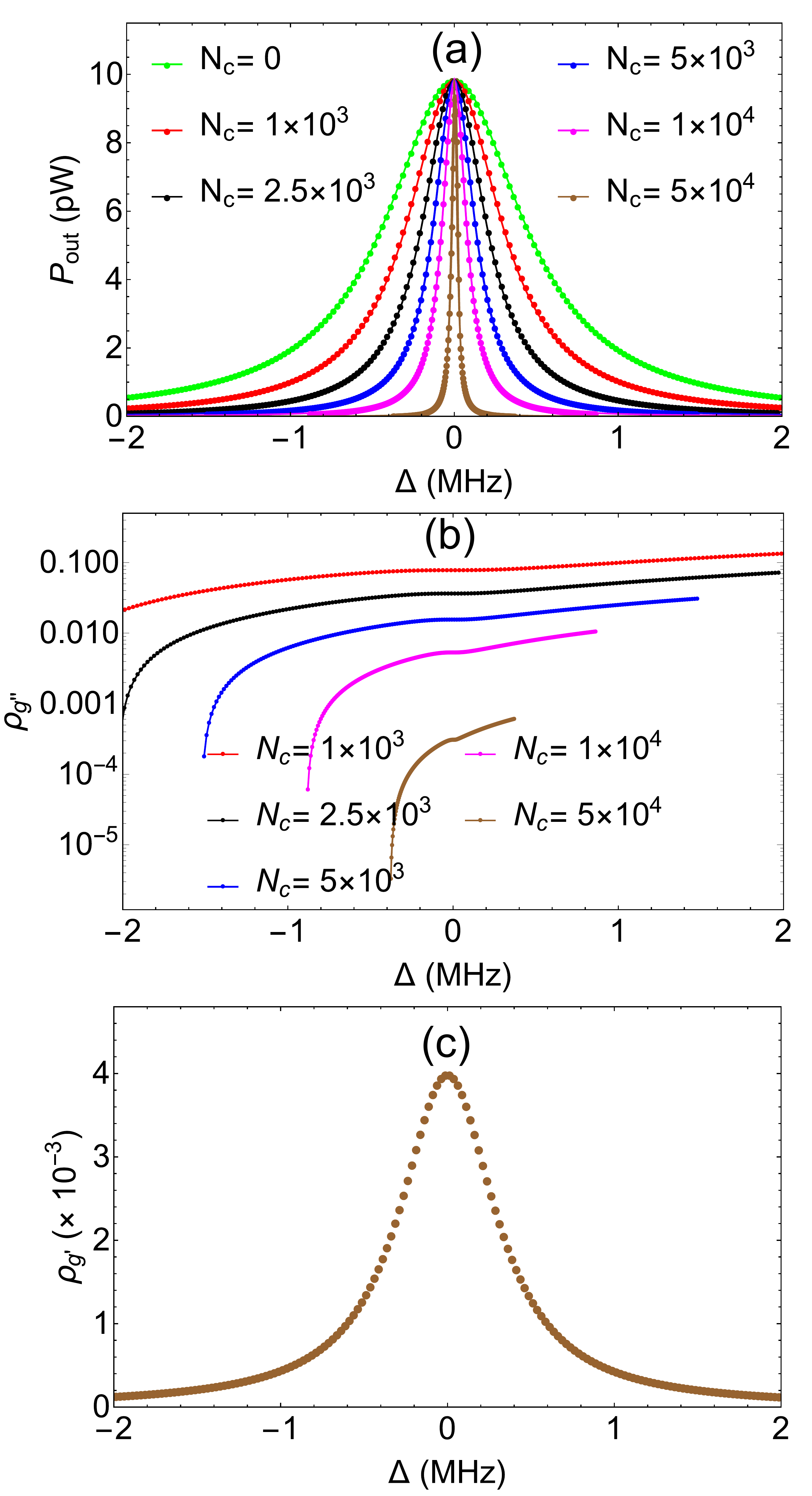}
	\caption{(a) The output of the cavity-atom system showing the EIT peak for various atom numbers. Dots show the results of time-dependent numerical simulation of the probe laser scan and the lines are obtained using Eqn.~\ref{fwhm_eit_simple}. (b) The probability for the atom to go into the dark ground state $\ket{\text{g}''}$. (c) The probability for the atom to be in ground state $\ket{\text{g}'}$ during the detection scan for $N_\text{c} = 1\times10^3$. At the start of the scan, all atoms are assumed to be in the state $\ket{\text{g}}$.} 
	\label{EIT_scan}
\end{figure} 

For such a scenario, the EIT peaks obtained by performing a numerical integration for different atom numbers are shown in Fig.~\ref{EIT_scan}. For Fig.~\ref{EIT_scan}, the parameters different from the numerical calculations of VRS (Fig.~\ref{3level_VRS_scan}) are, the power of light input to the cavity $P_\text{in} = 40$ pW, $\kappa_\text{t}/2\pi = 0.5$ MHz, $\Omega = 10$ MHz. Here, $\ket{\text{g}'}$ is the $\nu_\text{e} = 1, J_\text{e} = 0$ level of the $X^1\Sigma_\text{g}^+$ state, $\ket{\text{g}''}$ represents all other $X^1\Sigma_\text{g}^+$ levels.
Here, for $N_\text{c} = 1\times10^3$, $10\%$ atoms will to lost to the dark ground state per scan and for  $N_\text{c} = 5\times10^4$, $0.06\%$ atoms will to lost. For the analysis here, we have kept the scan duration to be 1 ms because it takes more time to reach steady state for detection using EIT. Additionally, at the end of the scan, very few atoms go to the state $\ket{\text{g}'}$ as can be seen from Fig.~\ref{EIT_scan}(c).  

%

This EIT detection is a significant improvement in terms of loss of molecules from the relevant detection transition compared to the VRS detection. For the same $N_\text{c}$, the loss rate per scan is two orders of magnitude better, and hence detection of a smaller number of molecules is possible. For such an EIT-based detection scheme, having smaller $\kappa_\text{t}$ is better because the EIT window is usually small. For example, if we keep the output photon flux and other parameters same, $\kappa_\text{t} = 2.5$ MHz will result in a loss of $80\%$ in the case of $N_\text{c} = 1\times10^3$ and a loss of $0.3\%$ atoms per scan in the case of  $N_\text{c} = 5\times10^4$. Although small $\kappa_\text{t}$ is desirable in this case, we cannot lower it arbitrarily because the linewidth of the probe laser will also require narrowing. Similarly to the case of detection using VRS, here too the ratio $\kappa_\text{r2}/\kappa_\text{t}$ should be on the higher side to avoid high intracavity photon number occupation.

\begin{figure}[h]
	\centering
	\includegraphics[width=7cm]{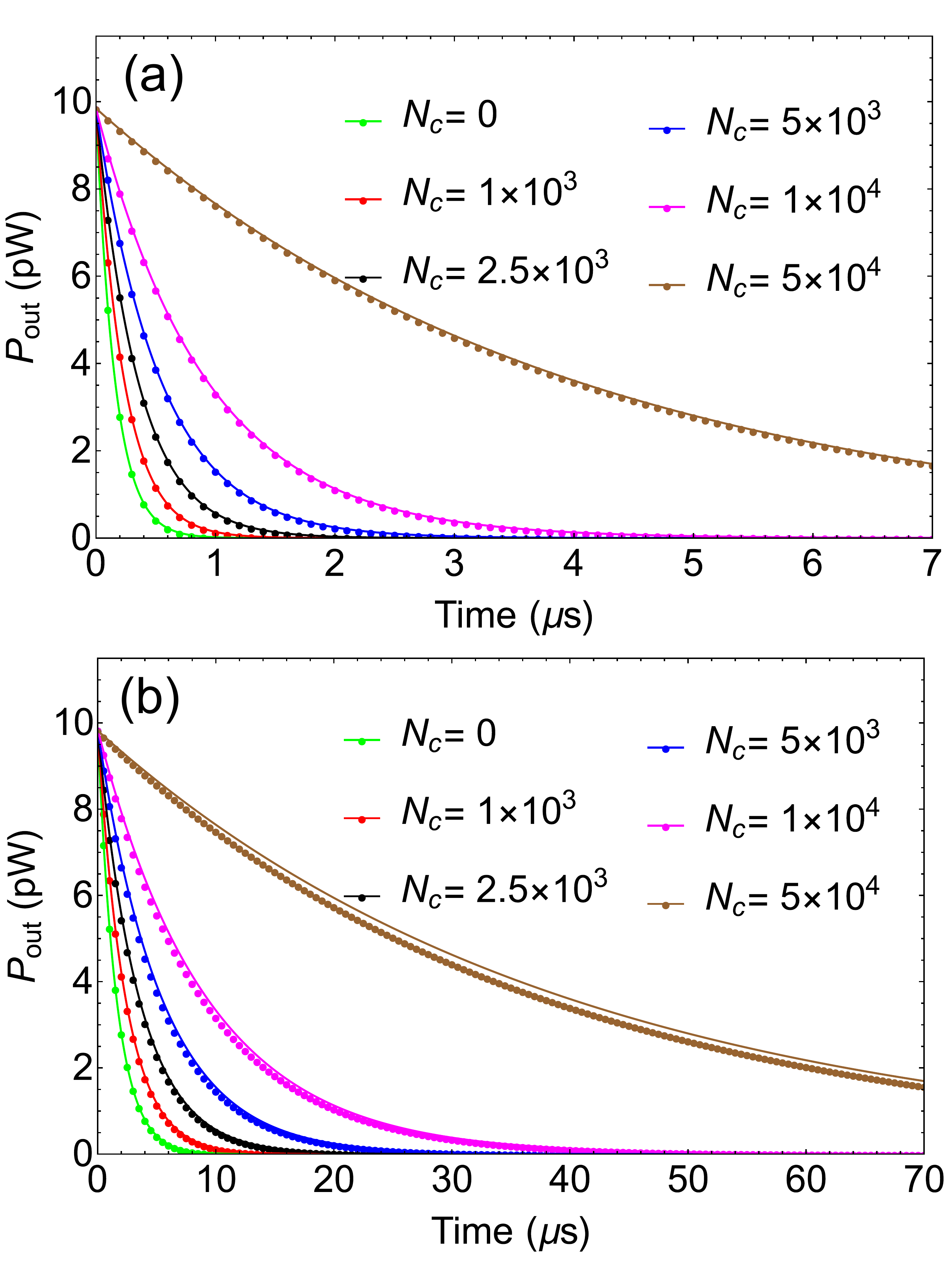}
	\caption{ The output of the cavity-atom system showing a slowing of cavity light decay for various atom numbers. (a) $\kappa_\text{t}/(2\pi) = 0.5$ MHz and (b) $\kappa_\text{t}/(2\pi) = 0.05$ MHz. All other parameters are same as in Fig.~\ref{EIT_scan}. Dots show the results of time-dependent numerical simulation of the decay and the lines are obtained using an exponential form, $I_0e^{-2d\times t}$ where $2d$ is taken from Eqn.~\ref{fwhm_eit_simple} and $t$ is the time variable.} 
	\label{slow_lgt}
\end{figure}

EIT is accompanied by the phenomenon of slowing of group velocity of the probe light~\cite{kasapiharris1995,kashscully1999,hau_1999, fleischhauer_electromagnetically_2005}. Hence, we expect an increase in trapping times for the photon inside the cavity as observed in previous experiments~\cite{shimizu_control_2002,zhang_slow_2008}. This can be exploited to detect the molecules placed inside the cavity. Below we explore a simple detection scheme, now in the time domain. The scheme is to set $\Delta_\text{pa} = \Delta_\text{pc} = \Delta_\text{ra} = 0$, to switch on the probe laser, and to allow the system to reach the steady state~\footnote{Time required to reach steady state is 0.1 ms for the atom number $5\times10^4$ and $\kappa_\text{t}$. It will be lower for smaller atom number} and obtain a constant light intensity output of the cavity. Then the probe laser is suddenly switched off and the decay of the cavity output light is observed. The results of numerical simulations are shown in figure~\ref{slow_lgt}.

A Lorentzian in frequency space implies an exponential decay in the time domain; we expect the decay curves to be exponential with the decay rates, $2d$ of Eqn.~\ref{fwhm_eit_simple}. This is indeed what is seen from the full numerical simulations in Fig.~\ref{slow_lgt}. For this detection scheme, a smaller value of $\kappa_\text{t}$ is very advantageous because the decay time and thus the observation time is higher, as seen in Fig.~\ref{slow_lgt}(b) for $\kappa_\text{t} = 0.05$ MHz. In Fig.~\ref{slow_lgt}(a) for $\kappa_\text{t} = 0.5$ MHz, the observation time is very short. Hence, very few photons will be collected during a single decay due to a finite dead time of an APD. However, multiple detection cycles can be performed, results of which can be added up. For example, the observation of one decay event in case of $\kappa_\text{t} = 0.5$ MHz gives a maximum of 1 photon detection per 50 ns, so for 10 detection cycles, we get a maximum of 10 detections per observation point if the total points are 20 for 1 $\mu$s decay. Similarly, in the case of $\kappa_\text{t} = 0.05$ MHz, we get a maximum of 10 detections per observation point if the total points are 20 for 10 $\mu$s decay~\footnote{This will require 500 ns integration window when counting photons.} if a maximum of 1 photon is detected per 50 ns. Comparing the two cases with respect to the total measurement time and the total photon flux, 10 decay events in case of $\kappa_\text{t} = 0.5$ MHz and 1 decay event in case of $\kappa_\text{t} = 0.05$ MHz give the same statistics for detection. 

For this detection scheme, the condition $\Delta_\text{pa} = \Delta_\text{ra}$ is always satisfied. Hence, we expect complete transparency and very little absorption of photons. For 10 detection cycles in case of  $\kappa_\text{t} = 0.5$ MHz, $0.09\%$ atoms are lost for $N_\text{c} = 1\times10^3$ and $6\times10^{-3}\%$ atoms are lost for $N_\text{c} = 5\times10^4$. In contrast, for single detection cycles in case of  $\kappa_\text{t} = 0.05$ MHz, $8\times10^{-3}\%$ atoms are lost for $N_\text{c} = 1\times10^3$ and $6\times10^{-4}\%$ atoms are lost for $N_\text{c} = 5\times10^4$. From above numbers, we can say that the detection involving lower $\kappa_\text{t}$ is more efficient. However, less than one atom is lost during each detection cycle in each case so the detection scheme with $\kappa_\text{t} = 0.5$ MHz results in very tiny loss of molecules from the ground state relevant for detection and it is not required to make the detection scheme less lossy. Less loss for this detection scheme implies more flexibility for the ratio $\kappa_\text{r2}/\kappa_\text{t}$. For example, if we fix the output photon flux, $\kappa_\text{t} = 0.5$ MHz, and other parameters, a low ratio of $\kappa_\text{r2}/\kappa_\text{t} = 0.01$ will result in loss of $6.5\%$ atoms in the case of $N_\text{c} = 1\times10^3$ and a loss of $0.4\%$ atoms in the case of $N_\text{c} = 5\times10^4$ MHz atoms for the 10 decay events as apposed to $65\%$ in case of detection through VRS. 

\section{Discussion}
When it comes to loss of molecules from the level used for detection, the detection of molecules through delay in decay times mentioned above is seen to be better than other detection schemes explored in this article. However, the condition $\Delta_\text{pa} = \Delta_\text{ra} = 0$ should always be satisfied, and any fluctuations around this condition will hinder the detection as the EIT effect is very sensitive near this condition as can be seen from narrow linewidths of Fig.~\ref{EIT_scan}. In the presence of such fluctuation higher value of $\kappa_\text{t}$ will be more advantageous due to large linewidths in this case. However, higher value of $\kappa_\text{t}$ will require multiple interrogation of the ensemble, which may present a problem for some experiments.   
Experimentally the fluctuations can be minimized by using stabilized probe and control lasers which are locked to a high finesse cavity~\cite{aikawa_narrow-linewidth_2011}. Availability of such a locking scheme makes such a detection process feasible.  

In the above analysis, we have ignored the hyperfine structure of the molecules. For some molecular states the hyperfine splitting will make the above calculation more complicated. But for singlet state the hyperfine splitting at zero magnetic field is typically of the order of few tens of kHz for alkali-metal diatomics~\cite{aldegunde2008,aldegunde2009,aldegunde2017,will2016,gregory2016controlling}. This is smaller than the cavity and atomic linewidths explored in this article and therefore should not create complications. This is supported experimentally	by the observation of the phenomena explored here, with atoms~\cite{wang_cavity-linewidth_2000}. For atoms the magnetic sublevels of the hyperfine levels are nearly degenerate and behave as a single level for the observation of VRS and EIT phenomenon. The other extreme where the hyperfine splitting is resolvable is also not a problem as this will just add more loss channels without affecting the detection process. In addition, for singlet molecular states the hyperfine splitting decreases when a high electric field is applied~\cite{aldegunde2008,will2016}. This can be exploited for species where the hyperfine structure creates problem.

\section{Conclusion}
In this article, we explored dispersion based non-destructive techniques with the help of numerical simulations and theoretical analysis for detecting molecules using a cavity. It is clear from the analysis that the large number of decay channels for molecules does not preclude cavity detection of molecules. Both VRS and EIT based arrangements were analysed and the parameters for molecule detection using each of these phenomena identified. The detection of molecules using the EIT feature is not just feasible, but also very efficient and with care can detect very small number of molecules. These techniques will be useful for not just detection of molecules but can be used to detect atoms/ions with multiple levels. A consequence of this study is that the need for a repumping laser is mitigated. Hence, the next logical step to advance these detection techniques will be to test them first on ultracold atoms as this is technically easier. Once this is demonstrated, the detection scheme can be extended to molecules. The analysis done here forms enables highly efficient exploration of cold, dilute molecular gases.    

\section*{Acknowledgement}
We thank Daniel Comparat for comments on the manuscript. R.S., S.A.R. and O.D. acknowledge support from the Indo-French Center for the promotion of Advanced Research-CEFIPRA Project No. 5404-1.

\bibliography{references.bib}{}
\bibliographystyle{unsrt}

\end{document}